\documentclass[prl,twocolumn, longbibliography]{revtex4-1}

\usepackage{bm}
\usepackage{amssymb}
\usepackage{graphicx}
\usepackage{epstopdf}
\usepackage{amsmath}
\usepackage{color}
\usepackage{microtype}
\usepackage{float}
\usepackage[english]{babel}
\usepackage{braket}
\usepackage{comment}

\newcommand{\BEQ}{\begin{equation}}
\newcommand{\EEQ}{\end{equation}}
\newcommand{\BEA}{\begin{eqnarray}}
\newcommand{\EEA}{\end{eqnarray}}

\newcommand{\sfa}{{\sf a}}
\newcommand{\sfb}{{\sf b}}

\newcommand{\iu}{\mathrm{i}}

\makeatletter
\newcommand\figcaption{\def\@captype{figure}\caption}
\makeatother

\begin{document}
\title{\sffamily \bfseries \Large The glassy random laser:
  replica symmetry breaking in the intensity fluctuations of emission
  spectra} \author{\sffamily \bfseries
  F. Antenucci$^{1,2}$, A. Crisanti$^{2,3}$ and L. Leuzzi$^{1,2}$}
\email{luca.leuzzi@cnr.it}
 \affiliation{ \small \sffamily $^1$
  NANOTEC-CNR, Institute of Nanotechnology, Soft and Living Matter
  Laboratory, Rome, Piazzale A. Moro 2, I-00185, Roma, Italy\\$^2$
  Dipartimento di Fisica, Universit\`a di Roma ``Sapienza'', Piazzale
  A. Moro 2, I-00185, Roma, Italy\\ $^3$ ISC-CNR, UOS {\it Sapienza},
  Piazzale A. Moro 2, I-00185, Roma, Italy }

\date{\today}

\begin{abstract}
{\sffamily \bfseries The behavior of a newly introduced overlap
  parameter is analyzed, measuring the correlation between intensity
  fluctuations of waves in random media in different physical regimes,
  with varying amount of disorder and non-linearity.  Its relationship
  is established to the standard Parisi overlap order parameter in
  replica theory for spin-glasses.  In the  complex spherical
  spin-glass model, describing the onset and behavior of random
  lasers, replica symmetry breaking in the intensity fluctuation
  overlap is shown to occur at high pumping or low temperature.  This
  order parameter identifies the laser transition in random media and
  describes its glassy nature in terms of emission spectra data, the
  only data so far accessible in random laser measurements. The
  theoretical analysis is, eventually, compared to recent intensity
  fluctuation overlap measurements demonstrating the validity of the
  theory and providing a straightforward interpretation of different
  spectral behaviors in different random lasers. }
\end{abstract}

\maketitle

\noindent {\sffamily \bfseries \large Introduction} \\ Light
amplification and propagation through random media have attracted much
attention in recent years, with present-day applications to, e.g.,
speckle-free imaging and biomedical diagnostics \cite{RedChoCao12},
chip-based spectrometers \cite{RedCao12,RedPopCao13,Redetal13}, laser
paints \cite{JohPan96} and cryptography \cite{Horetal13}.  Whatever
the amplifying medium, ordered or random, in a closed or in an open
cavity, two are the basic ingredients to produce laser in any
optically active system: \emph{amplification} and \emph{feedback}.  In
closed cavities the electromagnetic modes straightforwardly depend on
the cavity geometry.  In cavity-less random media, instead, some kind
of modes are established by spontaneous emission and
are localized in closed photonic trajectories by means of multiple
scattering.  Indeed, the phenomenon of {\em amplified spontaneous
  emission} (ASE) can occur even in systems without any optical
cavity, whose fluorescence spectrum is simply determined by the gain
curve of the active medium
\cite{Ambetal67,Ambetal67b,Ambetal68,Letokhov68,MarZolBri86,Gouetal93}.
When, because of an external pumping, the multiple-scattering feedback
process becomes strong, amplification by {\em stimulated} emission is
established in the random medium and we have a \emph{Random Laser}
(RL) \cite{WieLag96}.  The feedback is, here, associated to the
existence of well-defined long-lived localized modes, characterized by
a definite frequency and a spatial pattern of the electromagnetic
field inside the material.
Modes are expressed as {\em slow amplitude} contributions to the
electromagnetic field expansion in terms of spatial mode eigenvectors 
$\bm E_k(\bm r)$:
\begin{equation}
\label{eq:E_slow}
\bm E(\bm r,t)=\sum_{k} a_k(t) ~\bm E_k(\bm r)~e^{\imath \omega_k t} + \mbox{c.c.}
\end{equation}
The complex amplitudes $a_k(t)$ of these slow modes turn out to be the
fundamental degree of freedom in the statistical mechanical modeling
of interacting modes \cite{Antetal14,AntCriLeu14c}, while the
irregularity of their spatial profiles results into \emph{quenched}
disordered mode interactions.  By quenched we mean that the
interaction strengths are time independent \cite{Parisi06}, as it
occurs, in practice, when they change on time-scales much longer then
the typical amplification time-scales, possibly longer than the RL
lifetime itself.

At least in some random media, the RL action presents the peculiar
property of displaying strong non-trivial spectral fluctuations
\cite{SkiMay00,Anglos04,vanderMolen06,Leprietal07,Mujetal07,Fallert09},
i.e., narrow emission resonances appear to change frequency from one
excitation pulse of the pumping laser to another one
\cite{Cao98,Cao01a,Cao2001,Nakamura14}.
These will be termed \emph{shot-to-shot} fluctuations from now on.  If
in spectral fluctuation measurements the scattering particles and all
external experimental conditions are kept constant, these fluctuations
will only be due to the initial configuration of pre-pumping 
electromagnetic modes occurring because of spontaneous emissions.  

A connection to statistical mechanical models with quenched disordered
interaction, i.e., spin-glass models
\cite{MPVBook,Parisi06,Guerra09,Talagrand11}, has been recently
established
\cite{Antetal14,AntCriLeu14c,Angetal06,Angetal06b,Leuetal09,Conti11},
providing a new point of view on the shot-to-shot fluctuations
phenomenon.  The leading mechanism for the non-deterministic
activation of the modes is here identified with the frustration of the
disordered interactions and the consequent presence of a large number
of equivalent {\em states}. For {\em state} we mean a given ensemble
of activated mode configurations, specified by their own wavelengths,
phases and intensities, realized by very many emissions on time-scales
of the order of the duration of the shot, that is, of the RL life-time
itself.
The diverse spectral realizations are, thus, conjectured to correspond
to a glassy behavior consisting in many equivalent degenerate states
constituting the RL regime.  This glassy light regime is associated to
an effective thermodynamic phase where the tendency of the modes to
oscillate coherently in intensity is frustrated: in the language of
the replica theory \cite{MPVBook}, it corresponds to a phase where the
symmetry among equivalent replicas is spontaneously broken
\cite{Parisi79} and the overlap (i.e., the similarity) between the
configurations of the mode amplitudes display a nontrivial structure
\cite{AntCriLeu14c,Parisi83}.  Identical copies of the system show
different amplitude equilibrium configurations, as the ergodicity is
broken in many distinct states \cite{Parisi82}.

From an experimental point of view, the direct evaluation of the overlap
between complex amplitudes and its probability distribution, i.e.,
the standard order parameter of the theory, requires the measure of the mode phases in the coherent
regime. Such measure is not available so far, because of the low 
intensity of the RL emission (with respect to standard cavity lasers).
The lack of a direct experimental knowledge of the whole overlap
probability distribution is common, as well, to the original prototype
systems for which replica symmetry breaking (RSB) theory was first
developed, i.e., spin-glasses \cite{Edwards75,Sherrington75}, and
also to structural glasses, one of the fields of major application of
the theory
\cite{MezPar99a,ParZam10,Caltagirone12,Franz12,Charbonneau14}.

An experimental validation of such {\em random-glassy} laser
connection, and, particularly, of the RSB predicted by the theory,
has, nevertheless, recently been put forward in
Ref. [\onlinecite{Ghofraniha15}], measuring the overlap
between \emph{intensity fluctuations} of the spectral emission.
In the present work, we adopt a general model for cavity-less random
lasers, in which not only the mode phases \cite{Angetal06,Angetal06b,Leuetal09,Conti11} but the whole complex
amplitudes are considered as the fundamental degrees of freedom of the
problem \cite{Antetal14,AntCriLeu14c}.  In this framework we are
able to demonstrate that RSB occurring in the standard amplitude
overlap can, in principle, be observed in the intensity fluctuation
overlap (actually a coarse-graining of the former) and viceversa.  This
development provides a theoretical setting to explain existing
experimental results and to motivate similar
measurements in diverse RL systems.  Our approach also clarifies why
RSB is found only in RL's in which mode couplings can be considered
fixed (termed {\em quenched}) for all shots.
 In  liquid compounds, instead, as a TiO$_2$
dispersion in Rhodamine B-ethylene glycol solution, no evidence for
RSB is found \cite{Ghofraniha15}.

%%%%%%
\vskip .5 cm
\noindent {\sffamily \bfseries \large The Complex Amplitude Model}
\\ The statistical approach we adopt is based on the hypothesis of
effective equilibrium.  The steady-state of a laser can be described as if
at equilibrium at an effective temperature linked to the pumping rate
of the source and to the true environment temperature (associated,
e.g., to the noise of the spontaneous emission).
In the mean-field limit, the photonic system is described in this framework
by the general Hamiltonian
\cite{Antetal14, AntCriLeu14c}
\begin{equation}
 \mathcal{H} = - \frac{1}{2} \sum_{j k}^{1,N} J_{j k} a_{j} a^\ast_{k} 
 - \frac{1}{4!} \sum_{j k l m}^{1,N} J_{j k l m} a_{j} a^\ast_{k} a_{l} a^\ast_{m} \, , 
\label{eq:effective_Hamiltonian}
\end{equation}
where the sums are unrestricted and $a_i$ are $N$ complex amplitude
variables subject to the global power constraint $ {\cal E}=\epsilon
N=\sum_k | a_k |^2 $.  The coupling strengths are here quenched
independent random variables with mean $J_0^{(2)}/N^{p-1}$ and
variance $p!~J^2_p/(2N^{p-1})$ ($p=2,4$), whose scaling with $N$
guarantee an extensive Hamiltonian and thermodynamic convergence.  For
large $N$, the corresponding probability distribution can be taken
Gaussian without loss of generality.  Let us also define the degree of
disorder $R_J=J_0/J$ and the pumping rate $\mathcal{P}
=\epsilon\sqrt{\beta J_0}$ with $ J_0=J_0^{(2)}+J_0^{(4)} $ and
$J=J_2+J_4$, and where $\beta$ is the inverse of the environment
temperature.

This model can be derived in a multimode laser theory for open and
irregular random resonators \cite{AntCriLeu14c}.  The openness of the
cavity can be encoded into the definition of the electromagnetic modes
using, e.g., the system-and-bath approach of
Ref. [\onlinecite{VivHac03}], in which the contributions of radiative
and localized modes are separated by Feshbach projection
\cite{Feshbach62} onto two orthogonal subspaces.  This leads to an
effective theory on the subspace of localized modes in which they
exchange a linear off-diagonal effective damping coupling
\cite{Hacetal02,VivHac03,VivHac04}.  In terms of the interaction
parameters, we also define the strength of the openness as the inverse
strength of the nonlinear interaction coupling with respect to the
off-diagonal linear coupling $\alpha= J_4/J=J_0^{(4)}/J_0 \, \in
[0,1]$.  In a closed cavity the linear dumping is absent and it
corresponds to $\alpha=1$.

In a standard semiclassical approach, the field is expressed in the
slow amplitude basis, equation (\ref{eq:E_slow}), where each mode
displays a determined frequency.  The lifetimes of these modes are
assumed to be much longer than the characteristic times of population
inversion and amplification processes, so that the atomic variables
can be adiabatically removed and result in an effective interaction
between the electromagnetic modes.
The nonlinear couplings are, indeed, nonzero only for the terms $a_{j}
a^\ast_{k} a_{l} a^\ast_{m}$ that meet the \emph{frequency matching
  condition} \cite{Haus00,Gordon02,AntIbaLeu14b},
\begin{equation}
| \omega_{j} - \omega_{k}  + \omega_{l} - \omega_{m} | \lesssim \gamma \, ,
\label{eq:fmc}
\end{equation}
 $\gamma$
being the finite linewidth of the modes.  

The mean-field approximation of the model equation
(\ref{eq:effective_Hamiltonian}) is exact when the probability
distribution of the couplings is the same for all the mode couples
$(j,k)$ and tetrads $(j,k,l,m)$.  This is true, e.g., when mode
localizations scale with the volume occupied by the active medium and
their spectrum has a narrow-bandwidth around some given central
frequency $\omega_0$, i.e., $|\omega_j -\omega_0|<\gamma$, $\forall~
j$, so that the frequency matching condition, cf. equation
(\ref{eq:fmc}), always holds.

%%%%%%%

\vskip .5 cm
\noindent {\sffamily \bfseries \large Results}
\\
\noindent {\sffamily \bfseries The Random Laser Transition.}  Given
the quenched randomness of the $J$'s, any observable depends on the
particular realization of the disorder.  Thus, the relevant quantity
is the disorder averaged free energy $F=- {\overline{\ln Z_J}}/\beta$,
where the overline denotes the average over the distribution of
quenched disordered couplings.  This can be analytically evaluated
using the replica method \cite{Edwards75,MPVBook}, as reported in the
Methods.  In this procedure the evaluation of the relevant
thermodynamic quantities is achieved considering $n$ identical
replicas (i.e., copies) of the system that act as probes exploring the
multi-state phase space of the system. Further on, evaluating the distance
between the replicas in terms of their similarity, termed {\em
  overlap}, one can retrieve the physical overlaps of the
thermodynamic states \cite{Parisi83}.

In the complex amplitude
spherical model, equation (\ref{eq:effective_Hamiltonian}), the 
order parameter of the replica theory turns out to be given by
the overlap between amplitudes of replica $\sfa$ and replica $\sfb$:
\begin{equation}
\label{eq:Q}
 Q_{\sfa \sfb} = 
 \frac{1}{N \epsilon} \sum_{k=1}^N  a_k^{\sfa}\left(a_k^{\sfb}\right)^* 
\end{equation}
This Parisi overlap $Q$ identifies the onset of a RL regime at a given
critical value of the pumping, or, otherwise, at a critical
temperature at fixed pumping, since random lasing is known to occur
also decreasing temperature, besides increasing pumping
\cite{Wiersma01,Wiersma02,Nakamura10}. Any nontrivial structure of the
values taken by $Q_{\sfa \sfb}$ implies that identical copies of the
system, with the same interaction network and submitted to the same
thermodynamic conditions, show different sets of values for
microscopic observables at equilibrium and the ergodicity is broken in
distinct equivalent states.

To our knowledge, from an experimental point of view, no phase
correlation measurements, required for the evaluation of the complex
amplitudes $a_k^{\sfa}$ and, consequently, of $Q_{\sfa \sfb}$, is
available so far in random media.  Only magnitudes $|a_k|$ are
measured and not their phases $\phi_k=\arg(a_k)$. The experimental
reconstruction of the distribution of the values of equation
(\ref{eq:Q}) is, thus, unfeasible.

\noindent {\sffamily \bfseries Real Replicas.}  In recent experiments
\cite{Ghofraniha15}, shot-to-shot fluctuations of intensity spectra in
an amorphous solid RL, a functionalized thiophene-based oligomer named
thienyl-S,S-dioxide quinquethiophene (T5COx), are measured and
analyzed.  Since the sample remains under identical experimental
conditions shot after shot, $n$ different shots of RL emission
correspond to $n$ {\em real replicas} and one can measure the overlap
between \emph{intensity fluctuations} of two real replicas.  In these
experiments, the set of the activated modes emitting after the shot
${\sfa}=1,\ldots, n$, whose available coarse-grained degree of freedom
is the intensity $I_{\sfa}(k)=|a^{\sfa}_k|^2$, is observed to change
from shot to shot.

  When, during a single shot of the pumping source, the number of
  stimulated emission processes taking place is very large, the
  configurations of the mode dynamics can be considered as pertaining
  to a thermodynamic state.  In terms of the {\em photonic bomb}
  language of Letohkov \cite{Letokhov68}, e.g., this is a situation
  in which the typical amplification time is much shorter than the
  photons lifetime inside the medium, i.e., of the lifetime of
  stochastic resonators supporting the localized optical modes.  The
  possible observation of numerous different states from shot to shot
  is, consequently, an evidence of a thermodynamic phase described by
  a corrugated free energy landscape composed of many valleys
  separated by  barriers.

\noindent {\sffamily \bfseries Intensity Fluctuation Overlap (IFO).}
Having as only experimentally available degree of freedom the
intensities, one defines a suitable overlap based on their acquisition
in different shots.  To this aim, one first determines the average
emission spectrum $\bar I (k) \equiv 1/n \sum_{\sfa=1}^{n}
I_{\sfa}(k)$. Then, terming $\Delta_{\sfa}(k) \equiv I_{\sfa}(k)-\bar
I(k)$ the intensity fluctuation of shot $\sfa$ around the average
profile, one can define the overlap between the normalized intensity
fluctuation of shots $\sfa$ and $\sfb$ as \cite{Ghofraniha15}:
\begin{align}
\label{eq:overlap_neda}
 {\cal C}^{\rm exp}_{\sfa \sfb} \equiv 
 \frac{\sum_k \Delta_{\sfa}(k) \Delta_{\sfb} (k) }{\sqrt{\sum_k \Delta^2_{\sfa} (k)}
  \sqrt{\sum_k \Delta^2_{\sfb} (k)}},
\end{align}
where the index $k$ denotes now the frequency, i.e., the experimental
accessible equivalent of a mode index, depending on the spectral
resolution.  The overlap is measured between the fluctuations of
intensity, rather than the straight intensities, 
to exclude the effects due to the amplified spontaneous emission. 
From $n$ measured spectra one can calculate the $n (n -
1)/2$ values of the IFO $\mathcal C^{\rm exp}_{\sfa \sfb}$ and its distribution
$P_J(\mathcal C^{\rm exp})$.  
 Its average $P(\mathcal C^{\rm exp})={\overline {P_J(\mathcal
    C^{\rm exp})}}$ can be computed by repeated spectral measurements acquired
on different samples. By different samples we, actually, mean different
realizations of the microscopic disordered realization of scatterers
positions as faced by the incoming pumping light beam.  More
precisely, one can realize a different realization by turning the
material sample or, if the beam section is smaller than the random
medium, by illuminating  a different region with the pump laser spot.

If the variations of the normalization factors $\sqrt{\sum_k
  \Delta^2_{\sfa}(k)}$ in Eq. (\ref{eq:overlap_neda}) are 
  neglected with respect to fluctuations
$\Delta_{\sfa}(k)$,  in the $2+4$ complex amplitude
model given by equation (\ref{eq:effective_Hamiltonian}), 
the matrix 
\begin{align}
\label{eq:prescription}
 \mathcal{C}_{\sfa \sfb} \equiv & \frac{1}{8 N \epsilon^2}
 \sum_{k=1}^N \left[
  \langle |a^{\sfa}_k|^2  |a^{\sfb}_k|^2 \rangle
 - \langle |a^{\sfa}_k|^2 \rangle \langle |a^{\sfb}_k|^2 \rangle \right],
\end{align}
 is the model equivalent  of the IFO, up to an overall sign.
Indeed, equation (\ref{eq:prescription}), defined in the dominion $[0,1]$,
holds with the prescription that $P(C^{\rm exp}_{\sfa \sfb}= \mathcal C)$ corresponds to
$P(\mathcal{C}_{\sfa \sfb} =  |\mathcal C| )$. To compare with experimental results we will use symmetrized $P(\mathcal C)$,
$\mathcal C \in [-1:1]$, without any loss of generality.

\noindent {\sffamily \bfseries IFO vs. standard overlap relationship.}
The average in equation (\ref{eq:prescription}) can be
carried out using the replicated action derived in the Methods,
cf. equation (\ref{eq:S}). This leads to the following relationship
between the IFO and the standard order parameters:
\begin{align}
 \mathcal{C}_{\sfa\sfb}   = 
  Q_{\sfa\sfb}^2  - \frac{m^4}{4}
 \, , 
 \quad \sfa \neq \sfb \, ;
 \label{eq:C_Q2}
   \end{align}
where  $Q_{\sfa \sfb}$ is defined in equation (\ref{eq:Q}), 
and 
\begin{align*}
m \equiv m^\sigma_\sfa=&\frac{\sqrt{2}}{N}\sum_{k=1}^N
\mbox{Re}\left[a^\sfa_k\right]
\end{align*}
is the parameter of global coherence (cf. Methods). 
According to equation (\ref{eq:C_Q2}), if a RSB occurs in the
standard overlap $Q_{\sfa\sfb}$ it propagates with the same structure to the IFO
$\mathcal{C}_{\sfa\sfb}$.  We, thus, have a
theoretical well funded tool to detect RSB in experimental data.  We
stress that this analysis could have not been possible in 
$XY$ models with quenched amplitudes considered in previous works
\cite{Angetal06,Angetal06b,Leuetal09} because there
the intensities of the modes are kept fixed
during the mode dynamics.

In equation (\ref{eq:C_Q2}) we have considered the most general case
in which a high pumping regime can display both a global coherence
($m\neq 0$) and a multi-state non-trivial structure for the amplitude
configurations ($Q_{\sfa\sfb}\neq 0$). This mixing physically occurs
for a degree of disorder $R_J$ next to the tolerance value beyond
which standard mode locking (SML) breaks down, leaving place to random
lasing. This is displayed in the phase diagrams in the central panels
of the triptych figures \ref{fig:triptych1} and \ref{fig:triptych04}, as
the boundary lines between SML ($m\neq 0$) and purely random laser ($m=0$ but $Q_{\sfa\sfb}\neq 0$)
at large
${\mathcal P}$.

The low pumping regime is replica symmetric for any $R_J$, with $m=0$
and $Q_{\sfa\sfb}=0$ for $\sfa\neq\sfb$ \cite{Antetal14}, implying
a Dirac delta probability $P(\mathcal{C})$, peaked in zero, both in 
the incoherent wave (IW) and in the  phase locking wave (PLW) in figures 1 and 2 (cf. also Methods).

\noindent {\sffamily \bfseries Replica Symmetric standard mode-locking
  laser.}  For no or weak disorder at high pumping $\mathcal{P}$,
i.e., for a standard mode-locking laser in an ordered cavity
\cite{Haus00,Gordon02}, for every $\sfa\neq \sfb$ the relationship $
Q_{\sfa\sfb} = \pm m^2 / 2$, with $m \neq 0$, holds between the overlap and the
(replica independent) global coherence parameter.  In other words, the
ordered laser regime is replica symmetric, as well, and the
$P(\mathcal{C})$ is a Dirac delta function in zero, once again.
Remarkably, then, though in terms of the parameter $m$ the standard
mode locked regime is clearly different from the fluorescence regime,
and so is $P(Q)$, cf. left panels \textit{$a-d$} of figures
(\ref{fig:triptych1},\ref{fig:triptych04}) and
Refs. \cite{Antetal14,AntCriLeu14c}, {\em the IFO distribution does
  not change below and above the standard mode-locking transition}.
This has been observed in preliminary measurements on a Q-switched
pulsed Nd-Yag standard laser in Ref. \cite{Ghofraniha15}.  Indeed, the
overlap of the model in equation (\ref{eq:prescription}) is between
local fluctuations of intensity on different replicas, so a global
ordering is invariably taken away (cf. Methods).

\noindent {\sffamily \bfseries Onset of RSB across the random laser
  transition.}  For strong disorder, when the pumping increases above
threshold, $\mathcal{P}>\mathcal{P}_c$, the replica symmetry is broken
and the distribution of $\mathcal{C}_{\sfa\sfb}$ becomes nontrivial,
cf. panels \textit{f, g, h} in figures \ref{fig:triptych1} and
\ref{fig:triptych04}.

\noindent {\sffamily \bfseries Phase Diagrams and Overlap
  Distributions.}  Several scenarios are possible at the lasing
transitions, exemplified in the paradigmatic cases of figures
\ref{fig:triptych1} and \ref{fig:triptych04}.  In figure
\ref{fig:triptych1} we show the phase diagram and the behavior of the
Parisi overlap distribution $P(Q)$ and its relative symmetrized IFO
distribution $P(\mathcal C)$ in a closed cavity ($\alpha=1$) where
linear dumping is absent.  In figure \ref{fig:triptych04} the behavior
of $P(Q)$ and $P(\mathcal C)$ is shown across the laser threshold in
an open cavity ($\alpha=0.4 < 1$), where the linear dumping is
competing with non-linearity.  In both the closed and the open cavity
scenarios we illustrate two different critical regimes: the onset of
standard mode locking at low disorder and the transition to random
lasing for large $R_J$.

In a closed cavity situation, $\alpha=1$, $P(Q)$ is discontinuous at
the standard mode-locking laser transition, while $P(\mathcal C)$ is
unaffected.  In this case the transition itself is discontinuous in
the thermodynamic sense: the internal energy \cite{Gordon02,Antetal14}
and the coherence parameter $m$ (or the overlap $Q=\pm m^2/2$) are
discontinuous, see left panel of the closed cavity triptych in figure
\ref{fig:triptych1}.
Here the distribution $P(Q)$ has two values trivially
linked to the two possible values of the nonzero parameter, $m=\pm |m|$.
At the RL transition ($R_J=0.07$), alternatively, the $P(Q)$, and
similarly $P(\mathcal C)$, changes in a nontrivial way: two different
values, a zero and a nonzero one, are possible as the pumping is
increased.  In this situation the RL regime is one step RSB (1RSB) and
the transition is a so-called {\em random} first order (RFOT) in
glassy physics terming \cite{Parisi06}
\footnote{ In the RFOT scenario the static (ideal) glass transition is
  preceded by a glassy dynamic arrest (drawn as a dashed line in the
  central panel of figure \ref{fig:triptych1})
  \cite{Antetal14,AntCriLeu14c}: a photonic system in this case should
  show the typical two-step dynamical relaxation for the time
  correlation function of light modes, in the same universality class
  of the mode-coupling theory for structural glasses.  }: there is no
latent heat \cite{AntCriLeu14c}, yet a new value for the overlap
discontinuously appears at the transition, cf. right panel of figure
\ref{fig:triptych1}.

In the cavity-less scenario $\alpha=0.4$, instead, $P(Q)$ is
continuous at the ordered ML transition.  Indeed, a nonzero value for
$m$ increases continuously from zero as it can be observed looking at
the peaks of $P(Q)$ in panels \textit{a} and \textit{b} of figure
\ref{fig:triptych04}, where $Q=\pm m^2/2$.  As in the closed cavity
scenario, the $P(\mathcal C)$ of the ordered ML laser does not change
across the threshold.  At the onset of the RL regime, illustrated for
$R_J=1.1$, the $P(\mathcal C)$ is, instead, rather meaningful.  At and
just above the threshold, $P(\mathcal C)$ displays a continuous part
between the central peak in $\mathcal C=0$ and the two side peaks, as
displayed in panel \textit{$h$} of figure \ref{fig:triptych04}.  Here,
the transition is thermodynamically continuous with a RL regime that
is of the so-called full replica symmetry breaking (FRSB) kind,
associated with a free energy landscape composed by a fractal
hierarchy of valleys.
As the pumping increases, the regime becomes 1+FRSB, a combination of
1RSB and FRSB solutions, with both a continuous and a discontinuous
contribution to the probability distribution, cf. panel \textit{$g$}
in figure \ref{fig:triptych04}.  The continuous parts in the $P(Q)$
and $P(\mathcal C)$ depend on the influence of the off-diagonal
damping term in $J_{ij}$ in equation (\ref{eq:effective_Hamiltonian}).
For high enough pumping, well-above the threshold, the non-linear term
eventually becomes dominant \cite{AntCriLeu14c} and the solution,
cf. panel \textit{$h$} in figure \ref{fig:triptych04}, eventually
becomes 1RSB, as in the closed cavity case, cf. panels \textit{f, g,
  h} of figure \ref{fig:triptych1}.

In the RL experiment of Ref. [\onlinecite{Ghofraniha15}] the
distribution $P({\mathcal C}^{\rm exp})$, with ${\mathcal C}^{\rm
  exp}$ defined in equation (\ref{eq:overlap_neda}) is peaked in zero
at low pumping, while it becomes nontrivial with a triple and,
eventually, double peaked shape as the lasing threshold is overcome.
Although in comparison with the theoretical predictions for $N\to \infty$
the peaks of $P({\mathcal C}^{\rm exp})$ are smeared by noise effects and finite modes' number effects, 
in all regimes $P({\mathcal C}^{\rm exp}) \simeq P(-{\mathcal C}^{\exp})$.
In figure \ref{fig:the_exp} we display a comparison between the
analytic IFO distribution computed in our $2+4$ complex amplitude
spin-glass model, cf.  equation (\ref{eq:effective_Hamiltonian}), in
an open cavity and the experimental measurements of ${\mathcal C}^{\rm
  exp}$ in Ref. \cite{Ghofraniha15,Ghofraniha15cor}.

%%%%%%%%%%%%%%%%

\vskip .5cm 
\noindent {\sffamily \bfseries \large Discussion}

In this work we provide the theoretical analytical background for a
recently introduced order parameter \cite{Ghofraniha15} that allows to
probe the phenomenon known as replica symmetry breaking in random
lasers by means of experimentally accessible observables.  These are
shot-to-shot intensity fluctuations and the order parameter is the
distribution of the values of the overlap between intensity
fluctuations in different shots, as analytically defined in equation
(\ref{eq:prescription}).  Replica symmetry breaking is a known
property occurring in mean-field glasses, spin-glasses and hard
optimization problems.  The parameters of the theory have never been
measured, though, in any real system in these fields. In,
particular, no measurement of the Parisi overlap and its distribution
has been provided.  The only experimental measure, so far, of a
quantity possibly related to the standard RSB overlap been recently
carried out on a photonic system.  The system is an amplifying and
scattering random medium, the T5COx, \cite{Ghofraniha15} displaying
random lasing at high pumping. The parameter is the distribution of
the shot-to-shot intensity fluctuations overlap (IFO).  In the
framework of a recently introduced general statistical mechanics
theory of random photonic systems \cite{Antetal14}, in equation
(\ref{eq:matrix_C}) we give here an analytic proof of the relationship
between the IFO and the standard Parisi overlap, equation
(\ref{eq:Q}), and we provide measurable predictions for its behavior
in both ordered and random lasing systems below and above threshold
and, furthermore, both in the cases of discontinuous and continuous
transitions to the laser regime at the threshold.  In particular, the
transition in the probability IFO distribution of a random laser is
shown to be discontinuous (cf. figure \ref{fig:triptych1}) for closed
(or controllable, limited open) cavities while it becomes continuous
(cf. figure \ref{fig:triptych04}) for highly open cavity
nonlinear wave systems.  In the cavity-less case, where experimental
measurements are available in at least one case, in figure
\ref{fig:the_exp} we compare theoretical and experimental behavior of
the distributions of the IFO $\mathcal C$ from low to high pumping.

According to our results, a RSB is to be expected in random lasers
whose random configurations of scatterers are fixed, i.e.  {\em
  quenched}, for all analyzed shots.  That is, the dynamics of their
positions evolves on time-scales much longer than the whole experiment
and real replicas can be realized.  This is the experimental case of
the solid/powder samples of random lasers as GaAs powders
\cite{Noginov04,Nakamura10}, core-shell colloidal CdSe/ZnS quantum
dots \cite{Chen11}, ZnO powders \cite{Cao99} or pressurized pellets
\cite{Cao01}, and polymeric substances \cite{Lee99,Anni04,
  Ghofraniha15}. A notable counterexample might be porous gallium
phosphide (GaP) filled with a solution of Rhodamine and methanol
\cite{vanderMolen07,ElDardiry10}, in which spectral fluctuations are
reported to be minimal and the structure of the resonances, though
random, appears to be reproducible from shot to shot. IFO measurements
might yield, in this case, an ordered-like $P(\mathcal C)$, peaked in
zero both below and above threshold.

On the contrary, experiments on optically active random media whose
scatterer particles sensitively move between subsequent shots in a
single experiment, as in liquid solutions of Rhodamine and methanol
with particles of Titanium oxide \cite{Lawandy94}, Zinc oxide
\cite{Cao00}, pure Titania \cite{ElDardiry12}, or colloidal CdSe
quantum dots \cite{Augustine15} could establish no real replicas. Not
having the same quenched disorder in all shots might prevent the
observation of RSB. The overlap between copies of systems with
different realizations of the disordered couplings, indeed, is known to
be replica symmetric, as it has been shown in models with continuous
spherical variables \cite{Chenetal15}, of which our model in equation
(\ref{eq:effective_Hamiltonian}) is a generalization.  Similarly to
what happens in the ordered ML case, cf. left panels of figures
\ref{fig:triptych1} and \ref{fig:triptych04}, in that case the
occurrence of a trivial single peaked $P(\mathcal C)$ in $\mathcal
C=0$ is expected, both below and above ${\mathcal P}_c$. Such a
behavior has been observed in a liquid system of TiO$_2$ scattering
nano-particle suspensions in solution of Rhodamine and methanol
\cite{Ghofraniha15}.

Eventually, we would like to stress that, besides a rigorous
interpretation of recent experimental results for random lasers in
terms of replica theory, our results provide an exciting and easily
available test of spin-glass theory properties in continuous systems
without local magnitude constraints, as disordered photonic systems.

%%%%%%%%%%%%%%%%%%%%%%%%%%%%%%%%%%%%%%%%%%%%%%%%%%%%%%%%%%

\vskip .5 cm
\noindent {\sffamily \bfseries \large {Methods}}

\noindent {\sffamily \bfseries {Replica Theory and Order Parameters.}}
The most complicated system that we are considering in our theory is a
random system with disordered mode couplings that possibly display a
high pumping/low temperature phase with ergodicity breaking and the
occurrence of very many states.  By ``very many" we mean that their
number scales with the size of the system, i.e. the number $N$ of
optically active modes. These states are not related by any simple
relationship among them. That is, e.g., no simple $Z_2$ spin reversal
symmetry occurs between states, as in the Ising model, nor $SU(2)$
symmetry as in the XY model.  In the complex glassy case, to probe the
multi-state disordered thermodynamic phase, one, thus, considers $n$
copies of the system with exactly the same set of disordered
couplings, the $J$'s, and evaluates the disorder averaged partition
function $\overline{Z_J^n}$ of the replicated system.  A continuation
to real $n$ is, then, taken to evaluate 
\begin{equation}
-\beta F=
{\overline{
\ln
    Z_J}}=\lim_{n\to 0} 
\frac{\overline{Z_J^n}-1}{n}
\label{eq:F}
\end{equation}
As a result, $F$
is expressed as a functional in the replica space of the overlap
matrices
        \begin{eqnarray}
\label{eq:overlap_matrices_RL}
 Q_{\sfa \sfb} &=& 
 \frac{1}{N \epsilon} \sum_{k=1}^N  \text{Re}\bigl[a_k^{\sfa}\left(a_k^{\sfb}\right)^* \bigr], \\
 R_{\sfa \sfb} &=& 
  \frac{1}{N \epsilon} \sum_{k=1}^N  \text{Re}\bigl[a_k^{\sfa} \, a_k^{\sfb}\bigr],
\end{eqnarray}
${\sfa,\sfb}=1,\ldots,n$ being replica indexes.
The diagonal parts are 
\begin{equation*}
Q_{\sfa\sfa}=\frac{1}{N\epsilon} \sum_{k=1}^N\left|a_k^\sfa
\right|^2=1 \, ,
\end{equation*}
 by definition of the total power constraint, and 
\begin{equation}
R\equiv R_{\sfa\sfa}=\frac{1}{N\epsilon} \sum_{k=1}^N \left|a_k^\sfa\right|^2 e^{2\imath \phi_k} \, ,
\label{eq:Raa}
\end{equation}
yielding information about global phase coherence.  This parameter discriminates between the IW ($R=0$)
and the PLW ($R>0$) regimes (cf. figures 1 and 2), in which all the other parameters are identical \cite{Antetal14}

Alternatively, writing  $ a_k = \sqrt{\epsilon} \left( \sigma_k +\iu \tau_k \right) $, we can define the overlaps
of the real parts $\sigma$ or the imaginary parts $\tau$ of the complex amplitudes:
\begin{eqnarray}
\nonumber
 A_{\sfa\sfb} &\equiv& 
 Q_{\sfa \sfb} + R_{\sfa \sfb} =
 \frac{2}{N}\sum_{k=1}^N \sigma_k^{\sfa}  \sigma_k^{\sfb} 
 \, , \\
 \label{eq:definition_M}
 B_{\sfa\sfb} &\equiv& 
 Q_{\sfa \sfb} - R_{\sfa \sfb} =
 \frac{2}{N}\sum_{k=1}^N \tau_k^{\sfa}  \tau_k^{\sfb}  \, .
\end{eqnarray}

As the system size becomes sufficiently large, the free energy 
sample-to-sample fluctuations die out and the free energy, equation (\ref{eq:F}), 
becomes
independent of disorder, i.e., it is \emph{self- averaging}.  For
$N\to\infty$ the physical value of the matrices follows from the
extremization of the free energy functional. Because of the fact that
the number of independent elements of an overlap matrix is $n(n-1)/2$
(taken away the diagonal) in the limit $n\to 0$ the usual minimization
of the thermodynamic potential actually becomes a maximization in the
space of the overlap matrices. 
 To maximize $F$, a non-trivial Ansatz on the structure of $Q$ and $
 R$ is necessary.  Indeed, it can be shown \cite{Antetal14} that the
 most intuitive {\sl replica symmetric} solution, with $Q_{\sfa\sfb}$
 and $R_{\sfa\sfb}$ independent of $\sfa$ and $\sfb$, does not lead to
 a thermodynamically stable solution in the whole phase space: beyond
 the critical point, in the glassy phase, one must, hence, resort to
 spontaneous RSB.  Following the Parisi scheme \cite{MPVBook} the
 overlap matrices are, then, taken $\mathcal{R}$-step RSB matrix, with
 $\mathcal{R}\to \infty$ for a continuous full RSB (FRSB).  These are block matrices where the number of inner blocks $\mathcal{R}+1$ corresponds to the number of hierarchical levels in the multi-state phase space.

Depending on the value of $J_{2,4}$ the solution of the RL model
Eq. (\ref{eq:effective_Hamiltonian}) displays phases with different
RSB structures, ranging from 1RSB, FRSB to a combination of
discontinuous one step and continuous breaking (1+FRSB)
\cite{CriLeu13}.

\noindent {\sffamily \bfseries {Replicated Action.}}  In the replica
formalism, the averages of an observable $O[\{a\}]$ over the
equilibrium Gibbs distribution and over the quenched disorder can be
written as
\begin{align}
 \nonumber
 &\lim_{n \to 0} {\overline{
 Z_J^{n-1} \int \prod_{k=1}^N da^*_kda_k ~O[\{a\}]~e^{-\beta\mathcal{H}[\{a; J\}]}
 }}
 \\
 \nonumber
 =& \lim_{n \to 0} \int 
 \prod_{a=1}^n da^\star_\sfa d a_\sfa~
 O[\{ a \}]~e^{\mathcal{S}[a]} 
 \equiv \langle O[\{a\}]\rangle %_{\text{eff}}
\end{align}
where the   average $\langle \ldots \rangle$  {\em in the replica space} is evaluated with the replicated action
\begin{align}
\mathcal{S}=&  - \frac{1}{2} 
\sum_{\sfa,\sfb=1}^n \sigma_\sfa \left( \mathcal{A} \right)^{-1}_{\sfa\sfb}  \sigma_\sfb
 + \sum_{\sfa=1}^n h^\sigma_\sfa  \, \sigma_\sfa
\nonumber  \\
 & - \frac{1}{2} 
\sum_{\sfa,\sfb=1}^n \tau_\sfa \left( \mathcal{B} \right)^{-1}_{\sfa\sfb}  \tau_\sfb
 + \sum_{\sfa=1}^n h^\tau_\sfa  \, \tau_\sfa 
\label{eq:S}
\end{align}
Here we have introduced 
the matrices
\begin{align}
& \mathcal{A} \equiv  A -  \vec{m}^{\sigma} \otimes \vec{m}^\sigma, 
 & 
& \mathcal{B}\equiv B -  \vec{m}^{\tau} \otimes
\vec{m}^{\tau}.
\end{align}
and the effective fields
\begin{align}
  h^{\sigma,\tau} \equiv 2  m^{\sigma,\tau}  \left\{
   b_2 + 2 b_4 \left[\left(m^\sigma\right)^2 + \left(m^\tau\right)^2 \right]\right\}  
   \label{eq:h}
\end{align}
These are functions of the  {\em global coherence} parameters 
\begin{align}
\label{eq:def_mag}
 m_\sfa^{\sigma} &= \frac{\sqrt{2}}{N} \sum_{k=1}^N \sigma_k^\sfa \, ,
 & m_\sfa^{\tau} &= \frac{\sqrt{2}}{N} \sum_{k=1}^N \tau_k^\sfa,
\end{align}
analogous to the magnetization for spin models,
with coefficients 
  $b_2 = \beta J_0^{(2)}   \epsilon/4$, 
$ b_4 =   \beta J_0^{(4)} \epsilon^2/96 $.
After some algebra (see Ref. \cite{AntCriLeu14c} for details), the field $h^{\sigma,\tau}$ can be  expressed as
\begin{align}
  h^{\sigma} \equiv &
  \frac{m^{\sigma}}{\sum_{c} A_{ac}}, &
  h^{\tau} \equiv &
 \frac{m^{\tau}}{\sum_{c} B_{ac}}.
\label{eq:h_sp}
\end{align}
For weak disorder (low $R_J$) the global coherence $m^{\sigma,\tau}$ is non-zero above the lasing threshold
and must be included into the description. If disorder is
strong, though, in the frozen glassy phase, the global coherence is null:
$m^{\sigma,\tau}=0$.

Because it turns out that $\langle \sigma_\sfa \tau_\sfb \rangle=0$
\cite{Antetal14,AntCriLeu14c}, the integrals in the
$\sigma , \, \tau$ space factorize and the IFO $\mathcal{C}_{\sfa \sfb} $ defined in equation (\ref{eq:prescription})
takes the form
\begin{equation}
\label{eq:C_st} 
\mathcal{C}_{\sfa \sfb} = \frac{1}{8}\left(
 \langle  \sigma_{\sfa}^2  \sigma_{\sfb}^2 \rangle + \langle  \tau_{\sfa}^2  \tau_{\sfb}^2 \rangle  
-  \langle \sigma^2 \rangle^2 - \langle \tau^2 \rangle^2\right),
\end{equation}
where $\langle \sigma^2 \rangle = \langle \sigma_{\sfa}^2 \rangle$ and
$\langle \tau^2 \rangle = \langle \tau_{\sfa}^2 \rangle$, 
since single replica quantities do not depend on the replica index.

The replicated action $\mathcal{S}$ given in equation  (\ref{eq:S}) is quadratic. 
Thus, using the Wick's theorem,  for the averages in equation (\ref{eq:C_st}) we easily obtain
\begin{align}
 \mathcal{C}_{\sfa\sfb} = & \,\,
 \frac{  \mathcal{A}^2_{\sfa\sfb} }{4} 
+ \frac{\mathcal{A}_{\sfa\sfb} 
\left(m^\sigma\right)^2}{2}
+\frac{  \mathcal{B}^2_{\sfa\sfb} }{4} 
+\frac{\mathcal{B}_{\sfa\sfb} 
 \left(m^\tau\right)^2  }{2}
\label{eq:C_A}
\end{align}

Equation (\ref{eq:C_A}) can be further simplified since the physical
solutions of the model are either of the form $ Q_{\sfa\sfb} =
R_{\sfa\sfb} $ or $Q_{\sfa\sfb} = - R_{\sfa\sfb} $ ($\sfa \neq \sfb$).
Since the two solutions are equivalent, without loss of generality we
choose the first one, so that $m^\tau=0$ and equation (\ref{eq:C_A})
leads to
\begin{eqnarray}
 \mathcal{C}_{\sfa\sfb}  & =& 
  Q_{\sfa\sfb}^2  - \frac{m^4}{4}
 \, , 
 \quad \sfa \neq \sfb \, ;
 \label{eq:matrix_C}
   \\
 \mathcal{C}_{\sfa\sfa}   &=& 
  \frac{1+R^2}{2} - \frac{m^4}{4}
 \label{eq:matrix_Cd}
\end{eqnarray}
where  $Q_{\sfa \sfb}$ is defined in equation (\ref{eq:Q}), 
\begin{align}
R \equiv R_{\sfa\sfa}=&\frac{1}{\epsilon N}\sum_{k=1}^N \mbox{Re}\left[
(a^\sfa_k)^2\right]
\end{align}
is the parameter of partial coherence, cf. equation (\ref{eq:Raa}), and 
\begin{align}
m \equiv m^\sigma_\sfa=&\frac{\sqrt{2}}{N}\sum_{k=1}^N
\mbox{Re}\left[a^\sfa_k\right]
\end{align}
is the parameter of global coherence \cite{Antetal14,AntCriLeu14c}.
Equation (\ref{eq:matrix_C}) is one of our main results and is
discussed in the main text, cf. equation (\ref{eq:C_Q2}).


%\bibliography{FabBib_v2}

\vskip .5 cm
\noindent {\sffamily \bfseries {Acknowledgments.}} 
We thank Silvio Franz for
stimulating this work and for interesting discussions.   The research
leading to these results has received funding from the People
Programme (Marie Curie Actions) of the European Union's Seventh
Framework Programme FP7/2007-2013/ under REA grant agreement n.
290038, NETADIS project, from the European Research Council through
ERC grant agreement no. 247328 - CriPheRaSy project - and from the
Italian MIUR under the Basic Research Investigation Fund FIRB2008
program, grant No. RBFR08M3P4, and under the PRIN2010 program, grant
code 2010HXAW77-008.

\vskip 1cm
\noindent {\sffamily \bfseries {Author contributions statement.}}
F.A., A.C and L.L. conceived the theory, performed the computations
and wrote the manuscript.

\begin{widetext}

\begin{figure}[h!]
\begin{center}
\includegraphics[width= 0.99 \textwidth ]{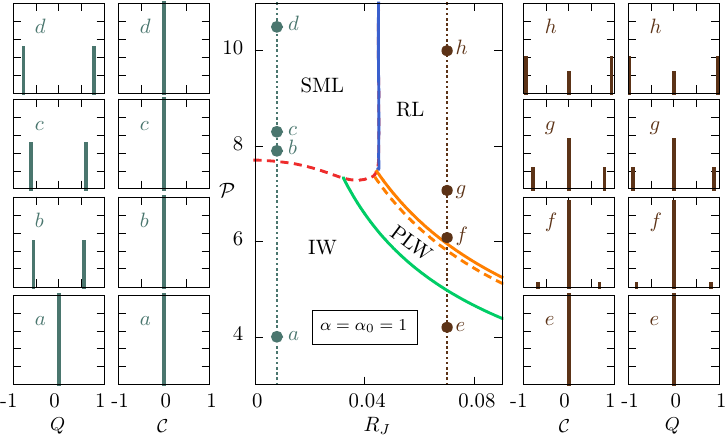}
\caption{Laser transition triptych in a closed cavity for varying
  disorder.  In the central panel the phase diagram $(\mathcal P,
  R_J)$ is displayed for a closed cavity (nonlinearity strength
  $\alpha=1$) in terms of the four possible optical regimes
  \cite{Antetal14,AntCriLeu14c}: incoherent wave (IW), standard mode
  locking (SML), phase locking wave (PLW) and random laser (RL).  Two
  pumping paths across the lasing tresholds are shown as dotted lines,
  at $R_J=0.01$ [$\mathcal{P} = 4.00 (a), 7.90 (b), 8.30 (c), 10.5
    (d)$] and $R_J=0.07$ [$\mathcal{P} = 4.20 (e), 6.08 (f), 7.07 (g),
    10.0 (h)$].  In the left panels $a$ to $d$ the behavior the
  distributions of IFO, $P(\mathcal C)$, and standard overlap, $P(Q)$,
  across the ordered ML laser threshold are reported. The transition
  is discontinuous in the standard Parisi distribution $P(Q)$, whereas
  $P(\mathcal C)$ is invariant.  In the right panels $e$ to $h$ the
  IFO and standard overlap distributions are shown for the RL
  transition: as $\mathcal P$ increases, we show
  that the low $\mathcal P$ solution is replica symmetric ($e$), while
  above threshold it becomes discontinuously 1RSB ($f$, $g$, $h$).  }
\label{fig:triptych1}
\end{center}
\end{figure}

\begin{figure}[h!]
\begin{center}
\includegraphics[width= 0.99 \textwidth ]{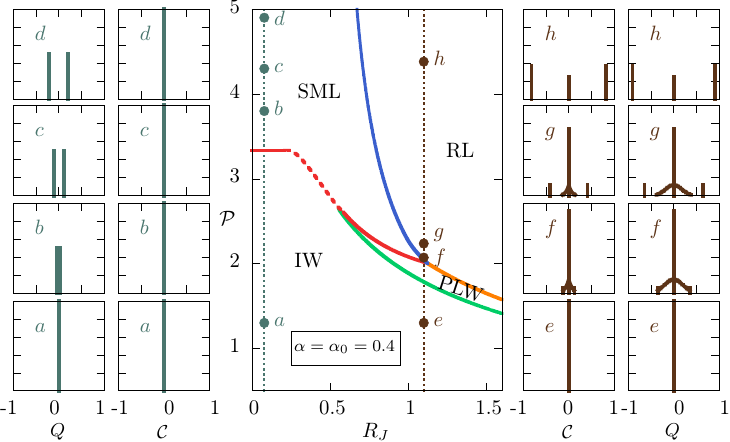}
\caption{Laser transition triptych in an open cavity for varying
  disorder.  In the central panel the phase diagram $(\mathcal P,
  R_J)$ is displayed for an open cavity (nonlinearity strength
  $\alpha=0.4$) in terms of the four possible optical regimes
  \cite{Antetal14,AntCriLeu14c}: incoherent wave (IW), standard mode
  locking (SML), phase locking wave (PLW) and random laser (RL).  Two
  pumping paths across the lasing tresholds are shown as dotted lines,
  at $R_J=0.07$ [$\mathcal{P} = 1.30 (a), 3.80 (b), 4.30 (c), 4.90
  (d)$] and $R_J=1.1$ [$\mathcal{P} = 1.30 (e), 2.05 (f), 2.35 (g),
  4.38 (h)$].  In the left panels $a$ to $d$ the behavior of IFO and
  standard overlap distributions across the ordered ML laser threshold
  are reported. The transition is now continuous in the order
  parameters $P(Q)$, while $P(\mathcal C)$ does not change below and
  above threshold.  In the right panels $e$ to $h$ the IFO and
  standard overlap distributions are shown for the RL transition. As
  $\mathcal P$ increases we show that the low optical power solution
  is replica symmetric ($e$), soon above threshold the solution is
  FRSB ($f$), further increasing $\mathcal P$ the solution becomes
  1+FRSB ($g$) and, eventually, for large pumping it is 1RSB
  ($h$). The transition is continuous in the order parameters
  $P(Q,\mathcal C)$.  }
\label{fig:triptych04}
\end{center}
\end{figure}

\begin{figure}[h!]
\begin{center}
\includegraphics[width= 0.9 \columnwidth ]{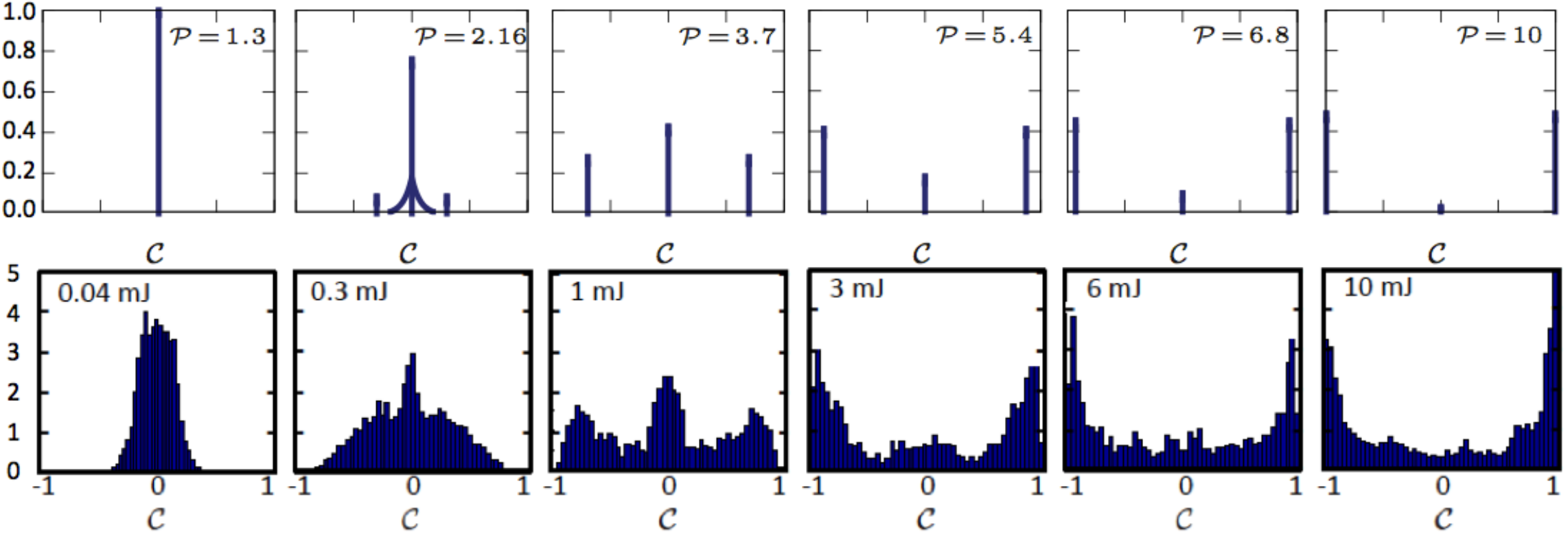}
\caption{Comparison between theory and experiments in a cavity less
  random laser.  In the top row we display the probability
  distributions of the IFO for $\alpha= 0.4$, when linear and
  nonlinear interactions are competing, $R_J=1.1$ and for increasing
  pumping. Vertical lines represent Dirac's deltas, whose height is
  the probability of the argument value. Different regimes are
  represented from fluorescence to large pumping random lasing. They
  are chosen along the dotted line in figure \ref{fig:triptych04} at
  $R_J=1.1$.  Form left to right the first distribution is at point
  $e$ in figure \ref{fig:triptych04}, the second between $f$ and $g$,
  the third one between $g$ and $h$ and the following above $h$. In
  the bottom row the same regimes are reproduced in the IFO
  distribution experimentally measured and reported in
  Ref. \cite{Ghofraniha15,Ghofraniha15cor} in an amorphous solid
  oligomeric random laser, T5COx.}
 \label{fig:the_exp}
\end{center}
\end{figure}

\end{widetext}

\end{document}